\documentclass{article}

\usepackage{PRIMEarxiv}

\usepackage[utf8]{inputenc} 
\usepackage[T1]{fontenc}    
\usepackage{hyperref}       
\usepackage{url}            
\usepackage{booktabs}       
\usepackage{amsfonts}       
\usepackage{nicefrac}       
\usepackage{microtype}      
\usepackage{lipsum}
\usepackage{fancyhdr}       
\usepackage{graphicx}       
\graphicspath{{media/}}     
\usepackage{float}
\usepackage{amsmath}
\usepackage{booktabs}
\usepackage{caption}
\usepackage[usenames,dvipsnames]{color}
\usepackage{color}

\newcommand{\change}[1]{\textcolor{black}{#1}}

\title{Genome-wide Nucleotide-resolution Model of Single-strand Break Site Reveals Species Evolutionary Hierarchy}

\author{
  Sheng Xu \thanks{These authors contributed equally.} \\
  Department of Computer Science and Engineering \\
  The Chinese University of Hong Kong \\
  Hong Kong SAR\\
  \texttt{shengxu@link.cuhk.edu.hk} \\
   \And
  Junkang Wei \footnotemark[1] \  \thanks{To whom correspondence should be addressed. Junkang Wei: weijk@link.cuhk.edu.hk; Yu Li: liyu@cse.cuhk.edu.hk.}\\
  Department of Computer Science and Engineering \\
  The Chinese University of Hong Kong \\
  Hong Kong SAR\\
  \texttt{weijk@link.cuhk.edu.hk} \\
   \And
  Yu Li \footnotemark[2]\\
  Department of Computer Science and Engineering \\
  The Chinese University of Hong Kong \\
  Hong Kong SAR\\
  \texttt{liyu@cse.cuhk.edu.hk} \\
}

\begin{document}
\maketitle

\begin{abstract} 

Single-strand breaks (SSBs) are the major DNA damage in the genome arising spontaneously as the outcome of genotoxins and intermediates of DNA transactions. SSBs play a crucial role in various biological processes and show a non-random distribution in the genome. Several SSB detection approaches such as S1 END-seq and SSiNGLe-ILM emerged to characterize the genomic landscape of SSB with nucleotide resolution. \change{However, these sequencing-based methods are costly and unfeasible 
for large-scale analysis of diverse species. Thus, we proposed the first computational approach, SSBlazer, which is an explainable and scalable deep learning framework for genome-wide nucleotide-resolution SSB site prediction. We demonstrated that SSBlazer can accurately predict SSB sites and sufficiently alleviate false positives by constructing an imbalanced dataset to simulate the realistic SSB distribution. The model interpretation analysis reveals that SSBlazer captures the pattern of individual CpG in genomic context and the motif of TGCC in the center region as critical features. Besides, SSBlazer is a lightweight model with robust cross-species generalization ability in the cross-species evaluation, which enables the large-scale genome-wide application in diverse species. Strikingly, the putative SSB genomic landscapes of 216 vertebrates reveal a negative correlation between SSB frequency and evolutionary hierarchy, suggesting that the genome tends to be integrity during evolution.}

\end{abstract}


\section{Introduction}

Single-strand breaks are represented as the most common DNA damage in the genome \cite{caldecott2008single}. 
The accumulation of SSBs caused by overstimulation of endogenous nucleases and defects in the DNA repair system would lead to genome instability, which has been implicated in multiple diseases such as cancer and neurological disorders \cite{tubbs2017endogenous}. 
Noting that the patterns of SSBs in the genome are non-randomly distributed, they are enriched in the regulatory element and exon region and vary in the differentiated cellular states.
Endogenous SSBs occur during conventional activities such as DNA replication, recombination and repair, and the major source is the oxidative attack by endogenous reactive oxygen species (ROS) \cite{pommier2003repair}. Besides, exogenous SSBs are the most frequent outcome of exposure to DNA-damaging agents such as ultraviolet (UV) and topoisomerase poison. In other words, SSBs can occur by the disintegration of the oxidized sugar directly or indirectly during DNA base-excision repair (BER) of oxidized bases, abasic sites and damaged bases in various pathways \cite{schreiber2002poly}. 

SSBs can impact cell fate in several pathways if they are not repaired immediately or appropriately. For example, it has been reported that SSBs can block transcription and induce cell death by inhibiting the progression of RNA polymerase \cite{kathe2004single}. Tyrosyl-DNA phosphodiesterase 1 (TDP1) and aprataxin (APTX) play vital roles in the SSB repair system, and several studies \cite{takashima2002mutation,clements2004ataxia} reveal that the absence of these key molecules has been found in various cancer and neurodegenerative diseases. 
The importance of SSBs has been underestimated since the dedicated cellular pathways that deal with every procedure of restoring SSBs, including SSB detection and repair procedures, have not been fully illustrated. Generally, defects in these pathways can induce genotoxic stress, embryonic lethality and several neurodegenerative diseases. 
Therefore, the description of the SSB landscape can provide a novel insight into these disease mechanisms and lay a foundation for the corresponding therapies \cite{rulten2013dna}. 

Studies \cite{kara2021genome,lensing2016dsbcapture,mehta2014sources} focusing on the double-strand breaks (DSBs) emerged to illustrate the DNA lesion repair system. The DSB-only detection approaches \cite{crosetto2013nucleotide,tsai2015guide,canela2016dna} have been developed for the genome-wide mapping of DSBs.
SSBs arguably pose less risk to cellular survival than DSBs. Generally, SSBs are regarded as the precursor of double-strand breaks in proliferating cells. By collision with replisomes, SSBs can stall replication forks and be transformed into DSBs, indirectly leading to genomic toxicity and cell death. 
However, The importance of SSBs has been underestimated since SSBs can also have a direct impact on disease progression. For example, Higo \textit{et al.} found that SSB-induced DNA damage is pivotal for the pathogenesis of pressure overload-induced heart failure. The accumulation of unrepaired SSBs would activate the DNA damage response system and increase the expression of inflammatory cytokines through the NF-$\kappa$B pathway in the mice model, leading to apoptotic cell death or cellular senescence \cite{higo2017dna}.

Thus, Mapping DNA lesions genome-wide is the key to understanding damage signals and the DNA repair procedure corresponding to the genome context and chromatin status. 
Recently, several high-throughput technologies of SSB detection, including S1 END-seq \cite{wu2021neuronal}, SSiNGLe-ILM \cite{cao2019novel} and GLOE-Seq \cite{sriramachandran2020genome}, have emerged, describing the genome-wide landscape of SSB. 
END-seq \cite{canela2016dna} is originally designed for double-strand breaks (DSBs) detection. By adding the single-strand-specific S1 nuclease, SSBs can be converted into DSBs. Moreover, dideoxynucleosides (ddN) are used to alleviate the SSB repair process. Therefore, S1 END-seq can detect SSB efficiently in a non-strand-specific manner. 
SSiNGLe-ILM \cite{cao2019novel} is the first next-generation sequencing (NGS)-based approach developed for SSB mapping and generalized for any lesion that can be converted into a nick with a free 3'-OH group. Based on the 3'-OH lesion, SSiNGLe-ILM can capture SSB sites in a strand-specific manner. 
These approaches for identifying particular types of lesions can map their positions with nucleotide-level resolution and accurately quantify the breaking possibility in the genome regions with a specific status. 
For example, these experimental methods can capture insightful genome SSB landscapes of various fundamental biological processes. They try to depict genome SSB landscapes in numerous cellular states such as health and disease status, distinct developmental stages, and the process in response to typical environmental stresses.
In a recent study, Wu \textit{et al.} \cite{wu2021neuronal} employed S1 END-seq to reveal that enhancer regions are the hotspots in neurons for DNA single-strand break repair by PARP1 and XRCC1-dependent mechanisms. These experimental approaches can successfully depict the SSB genome landscape. 
However, these \textit{in vivo/vitro} detection approaches are time-consuming and expensive. The experiment methods are based on high-throughput sequencing, indicating the high demand for sequencing equipment and limiting their extensive application. 
\change{Since recent studies \cite{cao2019novel,wu2021neuronal} have released genome-wide SSB distribution maps of multiple species and cell lines, the high-quality and well-curated datasets enable the construction of \textit{in silico} framework to depict the genome SSB landscape across the invalidated species and lead to further scientific discoveries.}

\begin{figure}[t]
  \centering
  \includegraphics[width=16cm]{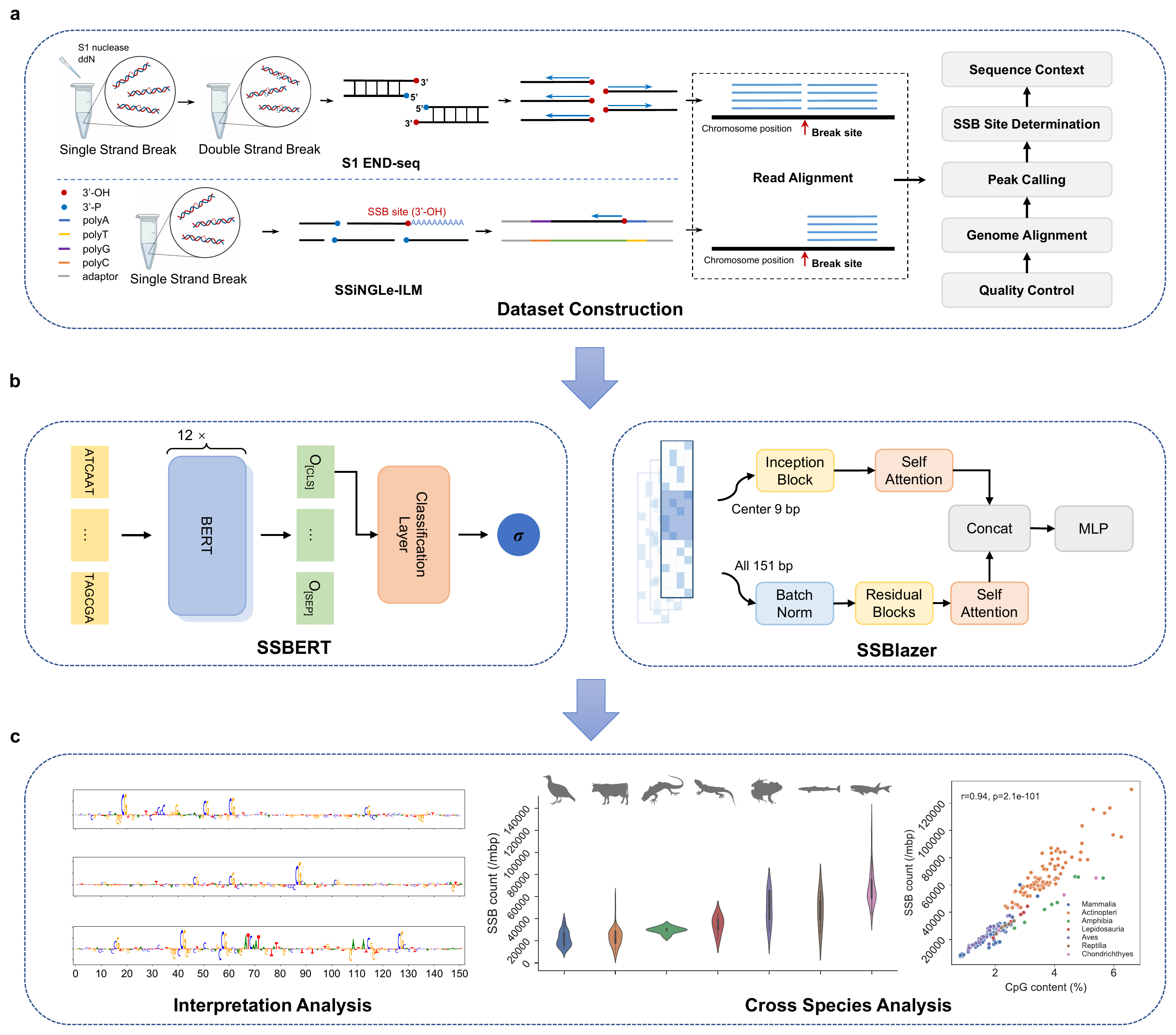}
  \caption{Overview of our work. \textbf{(a)} The experimental and bioinformatics pipeline for constructing the datasets from two different SSB detection methods (S1 END-seq and SSiNGLe-ILM). \textbf{(b)} The computational framework of SSBERT and SSBlazer. \textbf{(c)} The downstream analysis of the putative genome SSB landscape. For example, gradient-based model interpretation analysis reveals that the model can capture the specific motif pattern of the putative SSB sites. The cross-species analysis demonstrated that the number of SSB sites corresponded to the genome CpG content in various species.}
  \label{main_figure}
\end{figure}

\change{In this study, we proposed the first computational approach for genome-wide nucleotide-resolution SSB site prediction, namely SSBlazer, which is an explainable and scalable deep learning framework for genome-wide nucleotide-resolution SSB site prediction. 
We demonstrated that SSBlazer could accurately predict SSB sites. 
The construction of an imbalanced dataset simulates the actual scenario and sufficiently alleviates false positives.
Besides, SSBlazer is a lightweight model with robust cross-species generalization ability in the cross-species evaluation, enabling large-scale genome-wide application in diverse species. Strikingly, the putative SSB genomic landscapes on 216 vertebrates reveal a negative correlation between SSB frequency and evolutionary hierarchy. For example, fish, including \textit{Actinopteri} and \textit{Chondrichthyes}, usually share a high number of SSBs (26,860 to 130,088; 67,032 on average). At the same time, mammals have a relatively low SSB frequency (7,680 to 60,117; 23,398 on average). These results suggest that the genome tends to be integrity during evolution. 
The model interpretation analysis exhibits that SSBlazer captures the pattern of individual CpG in genomic contexts, which is consistent with the previous study \cite{wu2021neuronal} and provides novel potential insights into SSB occurrence mechanisms such as the break site motif of TGCC in the center region. 
The web server of SSBlazer is now available for the simplified application on \url{https://proj.cse.cuhk.edu.hk/aihlab/ssblazer/}, and the future version will expand the species and integrate genomic annotation features.}

\section{Materials and methods}

\subsection{Dataset construction}

\begin{table}[H]
\centering
\setlength{\abovecaptionskip}{0.5cm}
\caption{Datasets of single-strand break sites. Dataset I is reconstructed by the standard bioinformatics pipeline of END-seq analysis. Dataset II is provided in the original paper of SSiNGLe-ILM \cite{cao2019novel} and is established to evaluate generalization ability and cross-species performance.}
\resizebox{\linewidth}{!}{
\begin{tabular}{@{}ccccccc@{}}
\toprule
\multicolumn{1}{l}{\textbf{Index}} & \textbf{Cell lines} & \textbf{Methods} & \textbf{Validated SSB sites} & \textbf{Strand-specific} & \textbf{Assembly} & \textbf{Refrence} \\ \midrule
I   & i$^3$Neurons & S1 END-seq  & 63,386   & No  & hg19 & \cite{wu2021neuronal} \\ \midrule
II-A  & HeLa    & SSiNGLe-ILM & 2,331,388 & Yes & hg19 & \cite{cao2019novel} \\ \midrule
II-B & Neuro2A & SSiNGLe-ILM & 275,432  & Yes & mm10 & \cite{cao2019novel} \\ \bottomrule
\end{tabular}
}
\label{datasets-table}
\end{table}

\subsubsection{Data source}
\textbf{Dataset I.} The S1 END-seq data of human i$^3$Neurons cell line was collected from the study of Wu \textit{et al.} \cite{wu2021neuronal}. 
We applied the standard bioinformatics pipeline of END-seq for SSB site construction. First, the original sequencing data were downloaded from the Gene Expression Omnibus database (GSE167259), and the raw reads of ddN S1 END-seq sample were collected via SRA Run Selector. Following the standard pipeline of END-seq analysis, quality control was performed by fastqc (v.0.11.9) and trim galore (v.0.6.4) was applied for removing adapters and low-quality reads. The clean reads of END-seq were then aligned to the \textit{Homo sapiens} genome (hg19) via bowtie (v.1.1.2) and converted into a sorted bam file by samtools (v.1.14). After the genome alignment, peak calling was performed using MACS2 (v.2.2.7.1), and the summits of peaks were collected to generate SSB sites (positive sample). According to the summit coordinate, the upstream and downstream 75 bp sequence context was extracted from the \textit{Homo sapiens} genome (hg19), and the final input sequence length is 151 bp. 
Besides, we introduced a reverse complementary sequence of each positive sample for data augmentation since S1 END-seq is a non-strand-specific SSB detection approach. Finally, the positive dataset containing 126,772 sequences was constructed (Table \ref{datasets-table}). 

\textbf{Dataset II.} The SSiNGLe-ILM dataset was established to evaluate generalization ability and cross-species performance of the model, including Hela cell line (\textit{Homo sapiens}, II-A) and Nero2A cell line (\textit{Mus musculus}, II-B). Cao \textit{et al.} provided bed files of the integral SSB sites for Hela (GSM4126203) and Nero2A (GSM4126206) in the GEO database.
According to the SSB site coordinate, the upstream and downstream 75 bp sequence context was extracted from \textit{Homo sapiens} genome (hg19) and \textit{Mus musculus} genome (mm10), and the final input sequence is 151 bp. Since SSiNGLe-ILM provided a strand-specific insight, the sequence context of SSB sites was extracted based on the specific strand without adding reverse complementary sequences (Table \ref{datasets-table}). 

\subsubsection{Imbalanced dataset}
Negative set construction usually determines model practicality in scientific application. To simulate the real scenario where the SSB sites are rarely distributed in the genome, we introduced the imbalanced dataset (negative samples are more than positives) instead of the traditional balanced dataset to force the model to distinguish the genuine SSB sites. 
Here, we present the imbalance ratio \(Q\). Given \(N\) positive samples, the number of negative sequences is \(N\times Q\). We constructed three datasets for the positive sample using \(Q=1,5\) and \(10\) to investigate the role of the imbalanced dataset in model performance. 
The negative sequences were randomly sampled from reference genomes (hg19 and mm10) with a length of 151 bp, excluding the positive SSB sites. 
After the combination of positive samples and negative samples, we integrated the whole dataset, where 80\% of the shuffled dataset was used for training, and the remaining 20\% of it was for testing.

\subsection{Model framework}

\subsubsection{Data encoding}

In SSBlazer, the sequence context is encoded as a 4 \(\times\) \textit{L} matrix using one-hot encoding, where each nucleotide in the sequence is converted into a binary vector, with each dimension corresponding to a nucleotide channel \textit{A, C, G} and \textit{T} for the following convolution operation. Formally, given a DNA sequence $s=(s_{1},s_{2},s_{3}, ...,s_{n})$ with $L$ nucleotides, the one-hot encoding matrix \(M\) is:

\begin{equation}\notag
M_{i,j}=
\begin{cases}
1& \text{if}\ s_i=D_j\\
0& \text{otherwise}
\end{cases}
\end{equation}

where \(s_i\) is the \(i^{th}\) nucleic acid of the sequence and \(D =[A, C, G, T]\).

Since SSBERT is a DNABERT-based language model pre-training on the human genome (GRCh38.p13), SSBERT takes \textit{k}-mer tokens as the inputs, where \textit{k}-mers are substrings with defined length \(k\) of a sequence. Given a sequence of length \(L\), the set of \textit{k}-mer tokens is constructed by dividing the sequencing with the stride of 1, and the number of \textit{k}-mers is \(L-k+1\). In this study, we used 6-mer representations for enlarging sequence context receptive recognition to improve model classification capability. 

\subsubsection{Model construction}

\begin{figure}
  \centering
  \includegraphics[width=16cm]{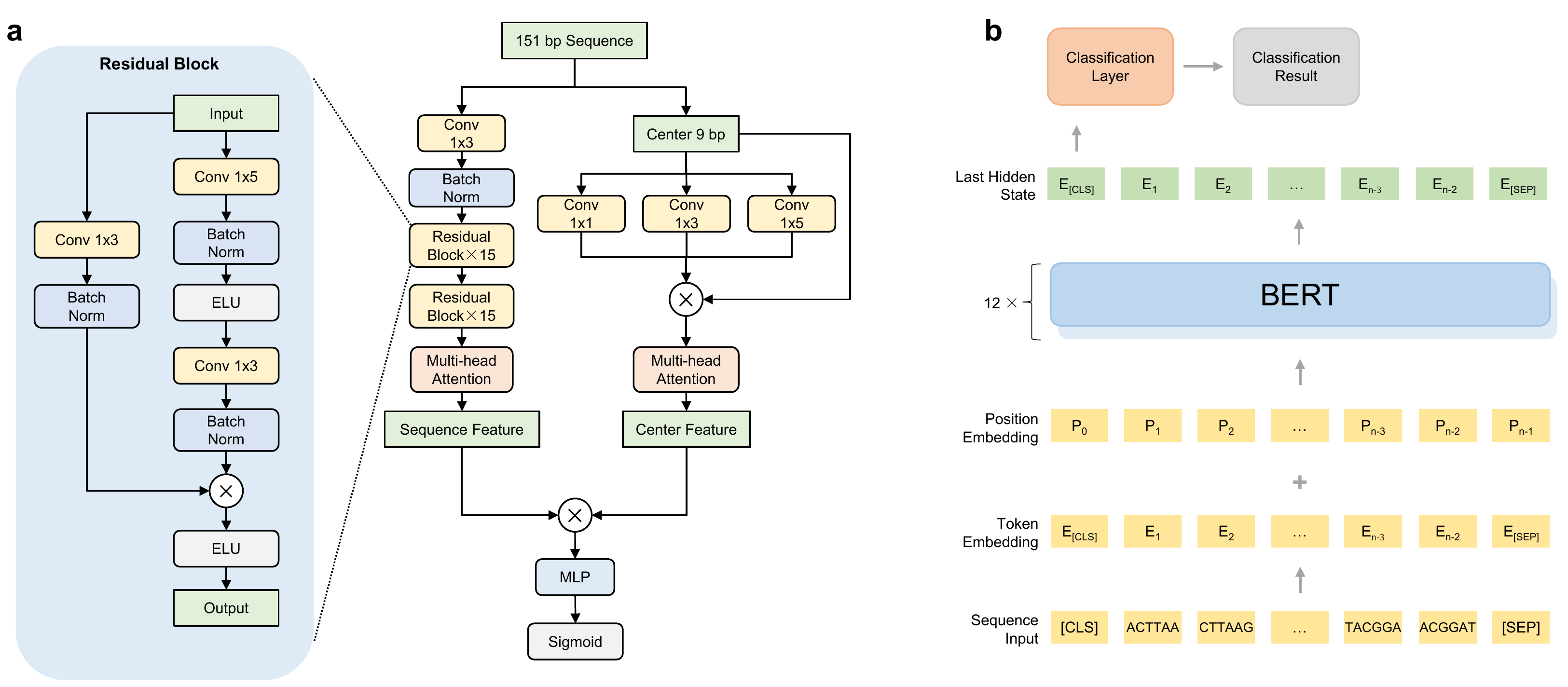}
  \caption{Overview of two proposed deep learning models. (\textbf{a}) SSBlazer uses two residual blocks to process the one-hot encoding matrix of the whole sequence. After processing by the residual network, we apply a multi-head self-attention layer (\(h=8, d_k=d_v=512\)) to extract long-range dependencies of the feature. The inception module captures the center feature, followed by a multi-head self-attention mechanism. The sequence context and center features are then concatenated and passed to the multilayer perception (MLP) for classification. (\textbf{b}) SSBERT is an attention-based Transformer model with 12 transformer blocks. The input sequence is segmented as \textit{k}-mer representations (\(k=6\)) as input tokens, which also contain two special tokens [CLS] and [SEP]. After the processing in the Transformer module, the first output (\(E_{[CLS]}\)) of the last hidden layer is regarded as the sequence feature and employed for further prediction.}
  \label{model}
\end{figure}

\textbf{SSBlazer.} The model takes one-hot representations of the 151 bp sequence as the input and processes the whole sequence and the center 9 bp separately. We use a residual-based neural network to obtain the vector representation of the input sequence for the feature extraction. Inspired by ResNet \cite{he2016deep}, SSBlazer (Figure \ref{model} a) starts with a 1 × 3 convolutional layer and follows by two sets of residual blocks with different convolution kernels. Each residual unit in the residual block consists of a 1 × 5 convolutional layer and a 1 × 3 convolutional layer. For the first 15 blocks, the number of filters \(N_{f1}\) is 16, and \(N_{f2}\) is 32 for the last 15 blocks. To alleviate the gradient vanishing problem, we apply the exponential linear unit (ELU) \cite{clevert2015fast} as the activation function instead of rectified linear unit (ReLU) \cite{nair2010rectified}. After a 1 × 5 average pooling, the sequence feature is then passed to a multi-head self-attention layer to capture long-range dependencies:

\begin{equation}\notag
\begin{aligned}
MultiHead(M) &= Concat(head_1,...,head_h)W^O \\
head_i &= Softmax\left(\frac{MW_i^Q {MW_i^K}^T}{\sqrt{d_k}}\right)MW_i^V
\end{aligned}
\end{equation}

where \(W^O \in \mathbb{R}^{hd_v \times d_{model}}\), \(W_i^Q \in \mathbb{R}^{d_{model} \times d_k}\), \(W_i^K \in \mathbb{R}^{d_{model} \times d_k}\) and \(W_i^V \in \mathbb{R}^{d_{model} \times d_v}\) are learned parameter matrices for projection.

In this model, we employ 8 heads (\(h=8\)) and set \(d_k=d_v=512\). \(d_{model}\) is depended on the input matrix (\(d_{model}\) = 64 in our cases). Nucleotides around the SSB site are considered an important pattern in several studies \cite{wu2021neuronal,kress2006active,fernandez2021epigenetic}. Thus, we collect the center sequence as the extra input to capture the specific pattern of center region. Due to the concise DNA sequence of center region, models with high complexity (\textit{e.g.}, ResNet) are tended to be overfitting. Therefore, we introduce a single-layer inception module \cite{szegedy2015going} to enrich the center information. The inception module contains three parallel convolutional layers, and the kernel dimensions are 1 × 1, 1 × 3 and 1 × 5 separately. The outputs of these layers are concatenated and combined with the original one-hot matrix of center sequence. Finally, the sequence context and the center feature are concatenated and fed into the multilayer perception (MLP) with two hidden layers for classification. Each fully connected layer is followed by a dropout with ratio of 0.5 and 0.3 separately to improve the generalization ability.

\textbf{SSBERT.} This model is a BERT-based language model, and all the parameters are initialized with those of DNABERT \cite{btab083}, which is a pre-trained language model on the human reference genome (GRCh38.p13). SSBERT (Figure \ref{model} b) uses \textit{k}-mer representations of the DNA sequence as input tokens, where tokens are first generated from the input sequence and then converted into numeral IDs by a predetermined mapping. In this stage, token IDs are used to obtain token embeddings. Segment embeddings and position embeddings are determined by the sequence and the position of each nucleotide. All these embeddings are vectors with a length of 768, and these vectors are then fed into an element-wise addition module to generate the final embeddings of the input sequence. Then, these embedding vectors are processed by 12 transformer blocks to obtain the final context feature, and each transformer block consists of 12 attention heads and 768 hidden units. The final embedding output (\(E_{[CLS]}\)) of the last hidden layer is regarded as the sequence feature and will be fed into the classification layer for prediction.

\subsubsection{Model interpretation}

To interpret the proposed model intuitively, we performed the integrated gradient analysis \cite{kokhlikyan2020captum} to illustrate how the individual nucleic acid influences the prediction outcome and determine potential motifs of predicted SSB sites. In contrast to extracting motifs from the putative SSB site context sequences, the integrated gradient analysis is a comprehensive gradient-based axiomatic attribution method that computes a contribution score for each input feature compared to a baseline. It can efficiently visualize the genomic pattern that the model captures. In this study, integrated gradient analysis emphasizes the contribution score of each nucleotide in the sequence containing an SSB site. Formally, \(x \in \mathbb{R}^n\) is the input sequence, \(x' \in \mathbb{R}^n\) is the baseline sequence, and \(F: x \in \mathbb{R}^n \rightarrow [0, 1]\) is the deep learning model, integrated gradients are calculated by accumulating the gradients \(\frac{\partial F(x)}{\partial x_i}\) along the path from the baseline input \(x'\) and the positive sample input \(x\), where \(x_i\) refers to the \(i^{th}\) feature:

\begin{equation}\notag
\begin{aligned}
IntegratedGrads_i(x) ::= (x_i-x'_i) \times \int_{\alpha =0}^{1} \frac{\partial F(x'+\alpha \times(x-x'))}{\partial x_i} d\alpha
\end{aligned}
\end{equation}

The baseline input \(x'\) should be a neutral input for the model, i.e.,  the prediction of it should be close to zero (\(F(x)\approx 0\)). In this research, we obtained the baseline sequence based on human GC content. Therefore, in the one-hot encoding matrix for the baseline sequence, each column is [0.3,0.2,0.2,0.3], representing the appearance probability of \textit{A, C, G} and \textit{T}.

\subsection{Construction of SSB site landscape in a species-wide manner}

After the construction of SSBlazer, we depicted vast landscapes of SSB sites species-wide. 
We first collected all the available genomes of vertebrates (216 reference genomes) from the Genbank and RefSeq database using python scripts (\url{https://github.com/kblin/ncbi-genome-download}). The scripts originally obtained 220 reference genomes. After the removal of the duplicates and classes that only contain one or two species, there are 216 species in total, including 91 species in \textit{Actinopteri}, 72 species in \textit{Mammalia}, 26 species in \textit{Aves}, 8 species in \textit{Amphibia}, 7 species in \textit{Reptilia}, 6 species in \textit{Lepidosauria} and 4 species in \textit{Chondrichthyes}. 
To remove the bias of various genome sizes, we employed normalized sampling, where we randomly generated one million sequences with 151 bp from each reference genome. We applied SSBlazer on these candidates for identifying SSB sites, and introduced SSB frequency to estimate the genome integrity. 
To visualize the evolutionary relationship among the 216 species, we employed iTOL (\url{https://itol.embl.de/}) to illustrate the evolutionary phylogenetic tree of these vertebrates based on species taxonomy ID. We integrated the SSB frequency heatmap into the evolutionary phylogenetic tree to discover the association between SSB frequency and evolutionary hierarchy.

\section{Results}
\subsection{SSBlazer accurately predicts SSB sites}

\begin{figure}
  \centering
  \includegraphics[width=16cm]{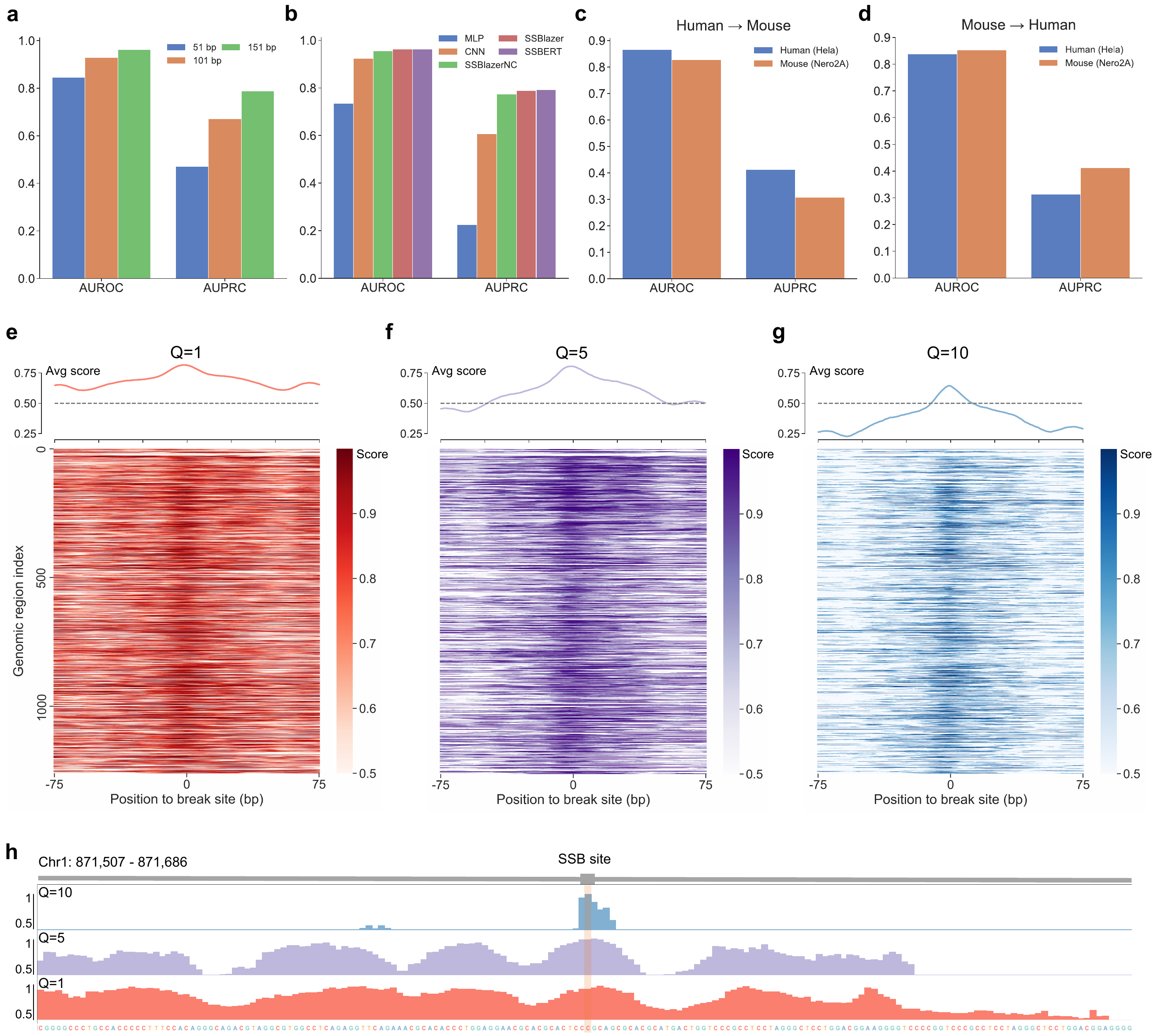}
  \caption{Model performance evaluation and outcome of imbalanced dataset introduction. 
  (\textbf{a}) \change{The AUROC and AUPRC value of different input sequence lengths. The model with a 151 bp sequence context shows superior performance among 151 bp, 101 bp and 51 bp.} 
  (\textbf{b}) The model performance comparison of the proposed models (AUROC and AUPRC). SSBlazerNC refers to SSBlazer without the center feature processing module. SSBlazer outperforms baseline machine learning models (MLP and CNN). 
  (\textbf{c, d}) Cross-species evaluation reveals that SSBlazer exhibits desirable cross-species generalization ability. SSBlzer was first trained the model on dataset II-A (\textit{Homo sapiens}) and evaluated the model performance on dataset II-B (\textit{Mus musculus}), and then had the reverse experiment (II-B for training, II-A for evaluation). A reverse experiment also reveals that SSBlazer achieves robust cross-species performance (II-A for training, II-B for testing). 
  (\textbf{e-g}) The profile heatmaps on 1,250 ground truth SSB sites represent the introduction of imbalanced datasets ($Q$=1, $Q$=5 and $Q$=10) on the 151 bp region around the putative SSB sites of the human genome hg19 chromosome 1. The signal-to-noise landscapes reveal that the introduction of imbalanced can sufficiently reduce false positives. 
  (\textbf{h}) Prediction scores on a ground truth SSB site region (Human chr1: 871,507 - 871,686) of $Q$=1 model (red), $Q$=5 model (purple) and $Q$=10 model (blue).
  The model trained in the traditional balanced dataset shows a high false positive rate in the flanking region. The model with the imbalanced dataset ($Q$= 10) has a significant narrow peak at the ground truth SSB site and a relatively low signal in the flanking region.}
  \label{performance}
\end{figure}

We evaluated the performance of SSBlazer on the S1 END-seq dataset, and the result suggests that the SSBlazer can accurately predict SSB sites. 
Besides, SSBlazer is a lightweight model, which achieved comparable performance to the BERT-based language model SSBERT with only 1.7\% parameter (SSBERT 110M, SSBlazer 1.9M) (Figure \ref{performance} b),  for both AUROC (SSBlazer = 0.9631, SSBERT = 0.9634) and AUPRC (SSBlazer = 0.7886, SSBERT = 0.7923) and laid the foundation of the large-scale downstream application. 
Since SSBlazer is the first computational approach for SSB site prediction, some baseline machine learning models such as MLP and convolutional neural network (CNN) can be proposed for performance comparison. Figure \ref{performance} b shows that SSBlazer significantly outperforms the other baseline machine learning models in both AUROC and AUPRC, indicating that SSBlazer can efficiently capture the pattern of SSB sites and has a superior capacity to distinguish the genuine SSB sites. 
For the ablation study, we investigated the contribution of the center feature. As shown in Figure \ref{performance} b, the removal of the center feature slightly decreased the model performance (AUROC by 0.72\% and AUPRC by 1.51\%), revealing the utility of the center context. 
Besides, We also investigated the influence of various lengths of SSB site context on the model. To this end, we constructed three variants of the S1 END-seq dataset with different sequence lengths (51 bp, 101 bp and 151 bp) to determine the optimal context length. Figure \ref{performance} a shows that the model with 151 bp context has superior performance on both AUROC and AUPRC, indicating that SSBlazer can capture long-range dependency and take advantage of dense context information to improve the model capacity.

\begin{table}[H]
\centering
\setlength{\abovecaptionskip}{0.5cm}
\caption{\change{Performance and computational complexity of SSBlazer and SSBERT. Each model was trained on 1 $\times$ NVIDIA RTX 3090 with 24 GB RAM, and the batch size is 128. MACs (Multiply–Accumulate Operations) was obtained by thop v.0.0.31 (\url{https://github.com/Lyken17/pytorch-OpCounter}). Training time is referred to the backpropagation time of the training set of S1 END-seq dataset, and inference time is the forward propagation time of the testing time. }}
\label{sclae}
\resizebox{\linewidth}{!}{
\begin{tabular}{@{}ccccccccc@{}}
\toprule
\textbf{Model} &
  \textbf{AUROC} &
  \textbf{$\uparrow \downarrow$} &
  \textbf{AUPRC} &
  \textbf{$\uparrow \downarrow$} &
  \textbf{MACs} &
  \textbf{Parameters} &
  \textbf{Training time (s)} &
  \textbf{Inference time (s)} \\ \midrule
SSBERT &
  0.9634 &
  0.0000 &
  0.7923 &
  0.0000 &
  1641.69G &
  110M &
  4216 &
  361 \\ \midrule
SSBlazer &
  0.9631 &
  0.0003$\downarrow$ &
  0.7886 &
  0.0037$\downarrow$ &
  16.33G &
  1.9M &
  851 &
  94 \\ \bottomrule
\end{tabular}
}
\end{table}

\subsection{\change{SSBlazer enables large-scale application with lightweight structure}}
\label{large-scale}
The massive demand for computing resources of the deep learning model usually prohibits large-scale downstream application. In this study, we intend to depict the SSB genomic landscape of various species. To this end, we demonstrated that SSBlazer achieved highly comparable performance with the BERT-based language model SSBERT but only with the demand of 1.7\% computing resources. 
We conducted the experiment on 1 $\times$ NVIDIA RTX 3090 GPU with 24 GB RAM using S1 END-seq dataset and determined the batch size as 128. 
The result shows that SSBlazer saves 99\% MACs and 98 \% parameters of SSBERT and maintains a comparable performance (AUROC reduced by 0.0003, AUPRC reduced by 0.0037). 
For further illustration, we applied SSBlazer on S1 END-seq dataset to obtain the training and inference time. For SSBlazer, the training time for a single epoch is 851 seconds, and the inference time is 94 seconds, while SSBERT takes 4,216 seconds for training and 361 seconds for testing.
However, the usage of GPU RAM for SSBlazer is only 3,401 MB with the insufficient batch size and is significantly lower than SSBERT's (20,005 MB), indicating that SSBlazer can further increase the batch size to take full advantage of computational resources and boost the training and inference process. These results suggest that SSBlazer is a lightweight model enabling large-scale parallel inference to depict SSB landscapes on diverse genomes.

\subsection{Imbalanced dataset reduces false positive prediction}
SSB sites are rarely distributed in the genome with a specific pattern. 
Constructing the dataset with an imbalanced positive-negative distribution is more rational to simulate the actual scenario, where the negative samples are more than the positive ones. 
To examine the role of the imbalance ratio $Q$, we trained the model with different datasets ($Q$=1, 5 and 10). 
We conducted a genome-wide enrichment analysis to identify the model's signal-to-noise status with various imbalance ratios. 
The profile heatmaps on 1,250 ground truth SSB sites (Figure \ref{performance} e-g) depict the signal-to-noise landscapes and reveal that all these models can successfully capture the genuine SSB sites in the center region.
However, the model trained in the balanced dataset ($Q$=1, Figure \ref{performance} e) shows a high false positive rate in the flanking region. In addition, the model training on the imbalanced dataset ($Q$=5, 10, Figure \ref{performance} f, g) has a relatively low signal on flanking sequence. Especially, the model with $Q$=10 only has a clear signal at the center region, indicating that the introduction of the imbalanced dataset can sufficiently reduce false positive prediction on flanking sequences, and the model training on the imbalanced dataset ($Q$=10) can precisely distinguish the genuine SSB site.
Specifically, the predicted scores on a ground truth SSB site (chr1: 871,507 - 871,686) are visualized by the integrative genomics viewer (IGV, Figure \ref{performance} h) and reveal that increasing imbalance ratio $Q$ can efficiently alleviate the false positive prediction. Besides, the model training on the imbalanced dataset ($Q$=10) presents a significant narrow peak around the actual SSB site, leading to an accurate enrichment of the predicted SSB site into the genuine site.
\change{These results reveal that SSBlazer training on an imbalanced dataset can efficiently reduce false positives, and the predicted SSB frequency is more corresponded to the reality genome-wide distribution, providing viability to describe the genome vulnerability.}

\subsection{SSBlazer exhibits robust cross-species generalization ability}
For further large-scale applications on diverse species, we demonstrated the cross-species generalization ability of SSBlazer. 
Since S1 END-seq dataset only covered the \textit{Homo sapiens} species, we integrated two datasets of \textit{Homo sapiens} and \textit{Mus musculus}, which were generated by the SSiNGLe-ILM approach for cross-species evaluation. 
Hence, We first trained the model on dataset II-A (\textit{Homo sapiens}) and evaluated the model performance on dataset II-B (\textit{Mus musculus}), and then performed a reverse experiment (II-B for training, II-A for evaluation).
Figure \ref{performance} c, d show that the model trained on the human dataset also achieved comparable performance on the mouse genome (AUROC: 0.8656, 0.8275; AUPRC: 0.4125, 0.3078), and the reverse experiment has a consistent conclusion (AUROC: 0.8374, 0.8523; AUPRC: 0.3138, 0.4131). \change{These observations indicate that SSBs in different species may share similar genomic patterns, and SSBlazer can successfully capture such patterns. Thus, SSBlazer has sufficient cross-species generalization ability and provides feasibility for further cross-species applications.}

\subsection{SSBlazer captures the pattern of CpG dinucleotides.}
\begin{figure}[htb]
  \centering
  \includegraphics[width=16cm]{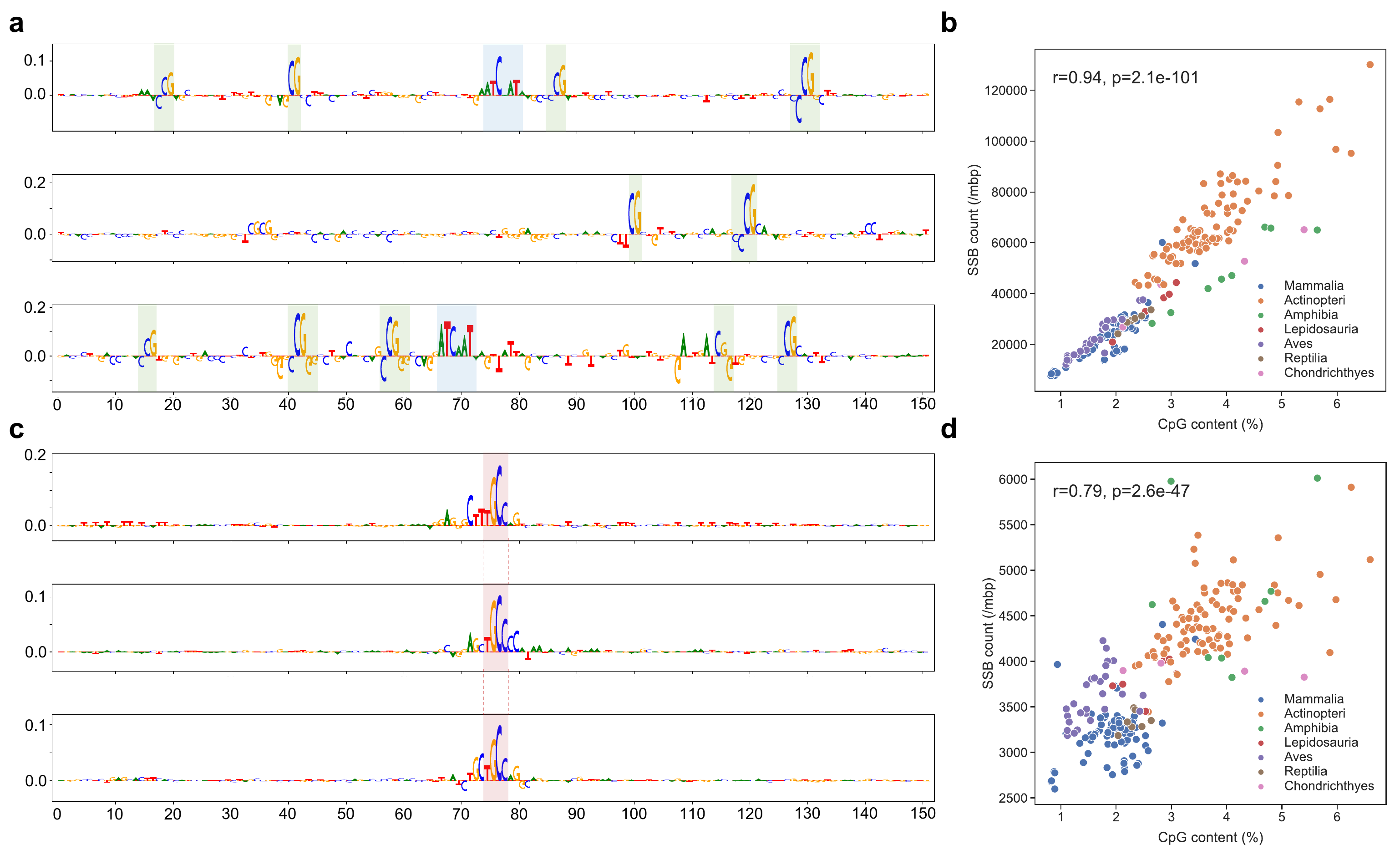}
  \caption{SSBlazer captures specific motif patterns of the SSB site context and determines the crucial role of CpG dinucleotide. (\textbf{a}) Integrated gradients analysis in S1 END-seq dataset reveals contribution scores of ground truth SSB site context. The individual CpG pattern is discontinuously distributed in the SSB site context, and the motif of CCG and CGG are disfavored in SSB site context. The motif of ATCAAT also arises in the SSB site context in S1 END-seq dataset. (\textbf{b, d}) Correlation analysis demonstrated the association between CpG content and predicted SSB counts in 216 species using Pearson correlation test (S1 END-seq: \textit{r} = \(0.94\), \textit{p} = \(2.1 \times e^{-101}\), SSiNGLe-ILM: \textit{r} = \(0.79\), \textit{p} = \(2.6 \times e^{-47}\)). (\textbf{c}) Contribution scores in the SSiNGLe-ILM dataset discover an explicit motif of TGCC in the center region of the SSB site sequence context.}
  \label{interpretation}
\end{figure}

To understand the SSB occurrence mechanism via the explainable model, we utilized the integrated gradients method by captum \cite{kokhlikyan2020captum} to determine which nucleotide decides the prediction outcome. 
We visualized the contribution scores on the individual putative sequences \cite{shrikumar2017learning} (Figure \ref{interpretation} a,c), and the direction of the nucleotide represents the contribution to the classification. If the nucleotide is above the x-axis, it positively contributes to the sequence being classified as a single-strand break site and \textit{vice versa}.
Among the sequences predicted to be SSB sites in S1 END-seq (Figure \ref{interpretation} a), CpG dinucleotides are observed to assign significant contribution scores. There are usually several individual CpG dinucleotides in an SSB site context, and such a pattern is consistent with the genome-wide landscape that SSBs are intended to locate at or near CpG dinucleotides \cite{wu2021neuronal}. 
Interestingly, the CpG pattern is discontinuously distributed, and the upstream cytosine and downstream guanine of CpG contribute negatively to putative positive SSB site prediction outcome, indicating that CCG and CGG trinucleotides are disfavoured in the SSB site sequence context. 
In addition, the motif of ATCAAT also arises in the SSB site context. 
Besides, we also conducted the interpretation analysis of the SSiNGLe-ILM dataset (Figure \ref{interpretation} c). An explicit motif of TGCC can be found in the center region of the SSB site context, indicating the distinct genomic pattern of that in S1 END-seq dataset, which can be explained by the bias from different SSB detection approaches and cellular states. 

As mentioned above, CpG dinucleotides are an important pattern in SSB site recognition. In addition, SSBlazer exhibits robust cross-species generalization ability. Thus, we performed the correlation analysis to determine the association between CpG content and putative SSB numbers in different species. 
The average content of CpG dinucleotides was obtained from normalized sampling to remove the bias from diverse genome sizes of different species. We randomly collected 1 million sequences for each species with a length of 100 bp from chromosome 1 and calculated the average CpG counts. 
It is worth noting that the correlation analysis in both S1 END-seq and SSiNGLe-ILM datasets shows a significant relationship between average CpG contents and SSB counts among various species. Figure \ref{interpretation} b, d reveal that the number of SSB increases in the species with higher CpG content and has a strong positive correlation with the CpG content (Figure \ref{interpretation} b, d, Pearson correlation test, S1 END-seq: \textit{r} = \(0.94\), \textit{p} = \(2.1 \times e^{-101}\), SSiNGLe-ILM: \textit{r} = \(0.79\), \textit{p} = \(2.6 \times e^{-47}\)), indicating that SSBlazer can sufficiently capture the pattern of CpG. 
\change{Besides, several studies \cite{kress2006active,fernandez2021epigenetic} have found that a possible source of DNA lesion is cytosine methylation/demethylation in the CpG region, which is consistent with the interpretation analysis. Thus, the explainable model of SSBlazer may provide novel potential insights into the SSB occurrence mechanism, such as the recognition motif of ATCAAT and the break site motif of TGCC.}

\begin{figure}[htb]
  \centering
  \includegraphics[width=16cm]{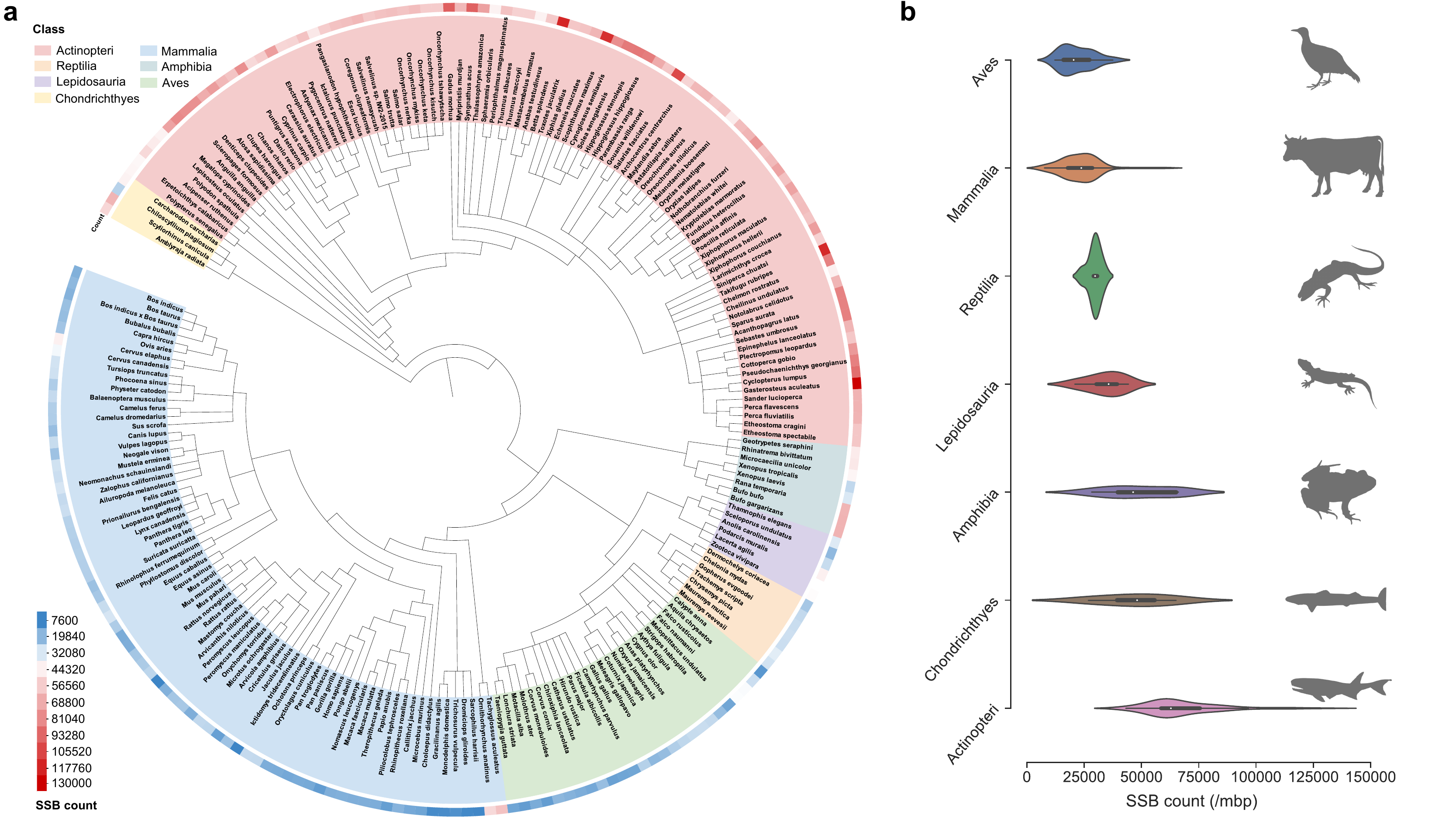}
  \caption{SSB frequency corresponds to species evolutionary hierarchy. (\textbf{a}) The phylogenetic diagram on 216 vertebrates indicates their evolutionary hierarchy, and the peripheral heatmap represents the number of SSBs. (\textbf{b}) Violin plots suggest order differences in SSB counts ranked by mean counts. Silhouettes of animals are downloaded from PhyloPic (\url{http://www.phylopic.org}).}
  \label{cross_species}
\end{figure}

\subsection{SSB frequency corresponds to species evolutionary hierarchy}

SSB frequency has been regarded as an implicit feature of genomic integrity \cite{zilio2021exploring}, which may associate with the evolutionary process differences in diverse species.
In order to clarify the relationship between genomic integrity and evolutionary hierarchy, we performed a large-scale SSB site analysis on 216 vertebrates. To this end, we first demonstrated that SSBlazer enables large-scale inference with a lightweight structure (section \ref{large-scale}) and generated realistic SSB genome distribution by introducing the imbalanced dataset. Furthermore, cross-species evaluation in human and mouse datasets reveals that SSBs in diverse species might share similar genomic patterns, and SSBlazer accurately captures such patterns and exhibits robust cross-species generalization ability.
Thus, we first collected all the available genomes of (216 non-redundant reference genomes) from the Genbank \cite{o2016reference} and RefSeq \cite{benson2012genbank}. We employed normalized sampling to remove the bias of various genome sizes. Specifically, we generated 1 million sample sequences with 151 bp from chromosome 1 of each reference genome. Then, SSBlazer was applied to a large number of sequences to identify the putative SSB sites and obtain the SSB frequency to estimate the genome integrity of each species. 

Strikingly, the frequency of SSB is highly variable among different species, ranging from 7,600 in thirteen-lined ground squirrel (\textit{Ictidomys tridecemlineatus}) to 130,088 in Alaskan stickleback (\textit{Gasterosteus aculeatus}) and a clear trend along the evolutionary process is observed. The phylogenetic diagram (Figure \ref{cross_species} a) reveals that \textit{Actinopteri} and \textit{Chondrichthyes} had relatively high levels of SSB frequency as well as \textit{Amphibia}. On the other hand, \textit{Reptilia}, \textit{Mammalia} and \textit{Aves} share fewer SSB frequencies. The violin plot (Figure \ref{cross_species} b) indicates the SSB counts distribution at the class level. Fish (\textit{Actinopteri} and \textit{Chondrichthyes}) and amphibians (\textit{Amphibia}) have higher SSB levels, suggesting that species in the aquatic environment are hard to maintain genome integrity. These results reveal that SSB frequency corresponds to the evolutionary hierarchy of species and may shed light on the development of evolution theory.

\begin{figure}[htb]
  \centering
  \includegraphics[width=16cm]{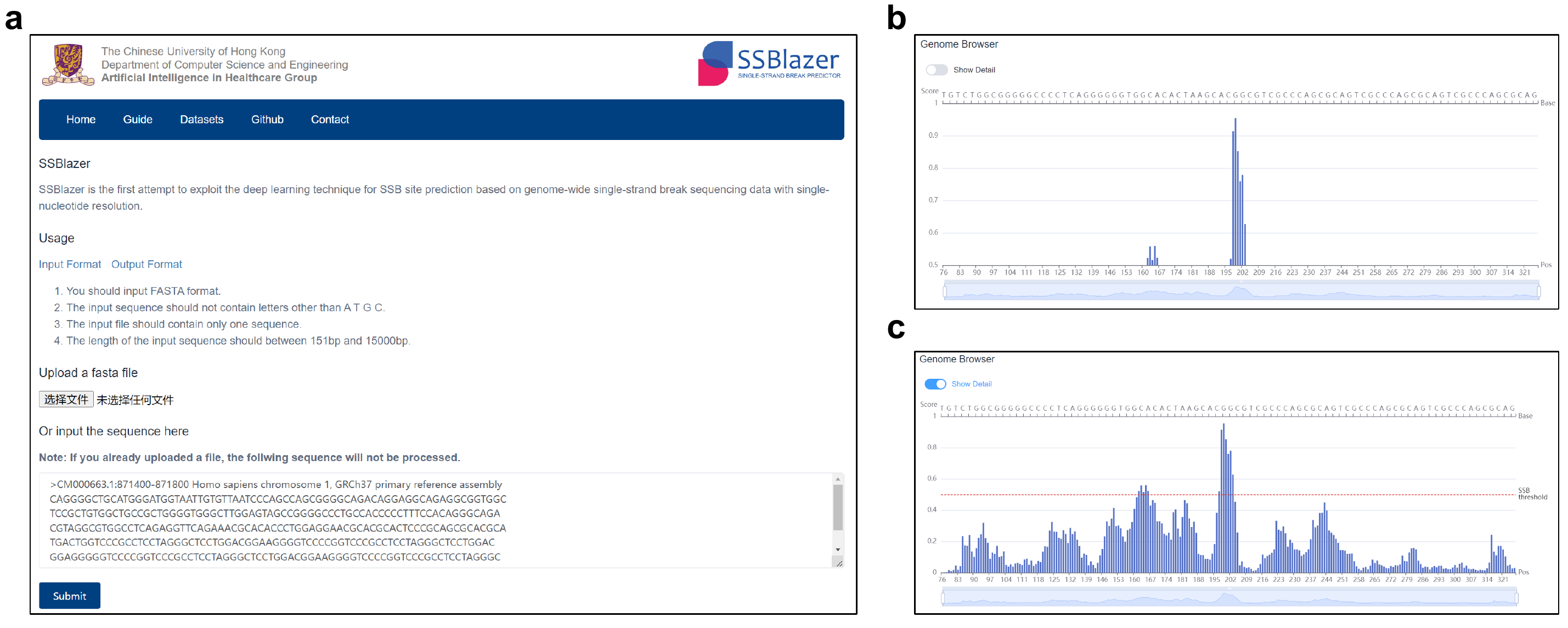}
  \caption{SSBlazer web server. (\textbf{a}) Screenshot of SSBlazer web server interface. The users can submit a single sequence up to 15,000bp in fasta file or input the DNA sequence directly into the text box. (\textbf{b, c}) The visualization of input sequence SSB score for each putative SSB site position. We also provide a toggle for SSB scores of the whole query sequence instead of the putative SSB sites.} 
  \label{webserver}
\end{figure}

\subsection{SSBlazer web server}

The accumulation of SSBs has been found in multiple diseases, while there is no available tool for SSB site prediction. 
\change{In this study, although we depicted vast landscapes of SSB sites for diverse species via SSB frequency, numerous SSB-related applications have not yet been explored. For example, we can employ SSB landscape within the target gene to evaluate the gene stability and single-nucleotide polymorphism (SNP). Besides, we can also elucidate the role of target site genome vulnerability in the CRISPR/Cas9 system and detect the association between SSB landscape and gene therapy outcome.}
Thus, for simplified usage, we established the web server of SSBlazer for intensive downstream applications. 
The users should provide a query sequence, and SSBlazer on the backend will process the sequence automatically and output the prediction score for each position of the query sequence (Figure \ref{webserver}). 
The output result will be saved in the bed format. We also integrate a genome browser to visualize the landscape of the SSB sites intuitively and provide a toggle for releasing the scores of the whole sequence instead of the putative SSB sites. 
Besides, the web server provided the processed dataset used in SSBlazer for further model construction improvement, and the new SSB dataset can easily be incorporated into SSBlazer to alleviate the detection approach bias. The web server of SSBlazer is now available at \url{https://proj.cse.cuhk.edu.hk/aihlab/ssblazer/}, and the future version will expand the species and genomic annotation features.

\section{Discussion}

High throughput SSB detection assays have emerged recently and provided approaches for describing SSBreakome generated by the direct action of several genotoxins or common intermediate products of DNA transactions. 
The landscapes of genome vulnerability derived from various cell lines can characterize the genome in various aspects such as evolutionary hierarchy, animal phenotype and cancer development. 
The NGS-based methods led to the discovery of SSBreakome, while these approaches are costly and highly demanded on sequencing equipment, indicating that it is hard to perform large-scale applications on diverse species. 
Besides, although the NGS-based methods shared a similar SSB detection strategy, the method bias exists in constructing the SSBreakome landscape. A unified and comprehensive SSB detection schema should be proposed in the future to alleviate the bias. The current version of SSBlazer is trained on the S1 END-seq dataset and the SSiNGLe-ILM dataset. SSBlazer is a portable and well-organized model, which can smoothly transfer to the unified dataset and other available SSB detection datasets (e.g. SSB-seq \cite{baranello2014dna}, GLOE-seq \cite{sriramachandran2020genome}) in order to improve SSBlazer generalization ability substantially.

\change{In this study, we proposed a deep learning-based framework to predict single-strand break sites by integrating the underlying SSB site context. To our knowledge, SSBlazer is the first attempt to exploit a computational model for SSB site prediction based on genome-wide single-strand break sequencing data with single-nucleotide resolution. 
We demonstrated that SSBlazer could accurately identify SSB sites and exhibit robust cross-species generalization ability. The introduction of the imbalanced dataset simulated the realistic SSB distribution in the genome and sufficiently reduced false positives. 
Notably, the interpretation analysis revealed that SSBlazer captures the pattern of individual CpG in the genomic context and the motif of TGCC in the center region as critical features and also provides the hypothesis for SSB occurrence mechanism exploration such as the recognition motif of ATCAAT in the SSB site context. Since SSBlazer is a lightweight model with robust cross-species generalization ability in cross-species evaluation, we conducted a large-scale genome-wide analysis on diverse species.
The putative SSB genomic landscapes on 216 vertebrates revealed a negative correlation between SSB frequency and evolutionary hierarchy.
A recent study \cite{lei2021genome} conducted genome-wide characterization of DNA microsatellite repeats in fish and found that the frequency of DNA microsatellite repeats plays a vital role in chromatin organization, recombination and DNA replication, indicating the abundance of DNA microsatellite repeats may be correlated with the high level of SSBs in \textit{Actinopteri} and \textit{Chondrichthyes}. 
Besides, a DSB detection approach called BLESS (Breaks Labelling, Enrichment on Streptavidin, and next-generation Sequencing) also found the enrichment of DSBs located in the microsatellite repeats region \cite{crosetto2013nucleotide}. 
These results demonstrated that such DNA lesions are non-randomly located in the genome and have specific distribution patterns. SSBlazer can distinguish the genomic pattern of SSB and may bring novel molecular insights into the physiological and pathological progress. }

Although SSBlazer achieved desirable performance, more efforts should be made to polish the SSBlazer in the future. 
Currently, SSBlazer only employed the DNA context as the input information, indicating that the model cannot precisely characterize the species genome and differentiated cellular states with the loss of genomic annotation such as genomic region, structure, and epigenetic information.
For example, the study \cite{cao2019novel} has revealed the enrichment of SSBs at the region of exons and other transcriptional regulatory elements such as CTCF binding sites. 
In addition, a DSB prediction model \cite{mourad2018predicting} demonstrated the usage of the DNA structural information such as minor groove width (MGW), propeller twist (ProT) at base-pair resolution, roll (Roll) and helix twist (HelT), which can efficiently improve the model performance. The result reveals the crucial role of DNA structural information in lesion recognition. 
\change{In CRISPR-induced DSB prediction \cite{chuai2018deepcrispr,lin2020crispr}, the epigenetic landscapes (\textit{e.g.} CTCF, Dnase, H3K4me4 and RRBS) are regarded as additional features of the model. The ablation experiment also reveals the importance of such chromatin status markers in the CRISPR-induced cleavage process. }
Thus, the future version of SSBlazer may include extra genomic information such as genomic region, structure and epigenetic information to characterize the vast and unique SSB genomic landscape in diverse species genomes, differentiated cellular states and distinct development stages. 

\change{The genome landscape of the SSB sites reveals the association between DNA lesions and various cellular conditions. Thus, landscapes of genome vulnerability derived from the various cell lines and phenotypes may contribute to previously unemployed insight for developing molecular biomarkers of disease diagnosis, ageing identification and gene therapy. 
Specifically, SSBs could lead to the accumulation of somatic mutations and transcription stalling in functional neuronal genes, contributing to neurological diseases \cite{el2007dna}, which have been reported by association with defective SSB repair systems. 
In addition, DNA lesions have been widely implicated in human ageing, and a recent study \cite{cao2019novel} emphasized the association between SSBreakome patterns and chronological age in humans. 
With the genome vulnerability landscape on the target site of gene therapy, the optimal gene editing approach can be designed (\textit{e.g.} high efficiency and fidelity sgRNA) to improve gene therapy outcomes. Besides, such SSB landscape may also offer insight into recognition and cleavage mechanism in the CRISPR/Cas9 system. }
SSBlazer achieved satisfying performance and exhibited strong cross-species generalization ability. The putative genomic landscape of the SSB site characterized by SSBlazer on diverse species may shed light on the mechanism exploration of ageing and complex diseases in various animal models. 

\section{Data availability} 
SSBlazer web sever is now available at \url{https://proj.cse.cuhk.edu.hk/aihlab/ssblazer/} for simplified usage and the source code of SSBlazer is available at \url{https://github.com/sxu99/ssblazer}. The dataset used in this study can be assessed at \url{https://proj.cse.cuhk.edu.hk/aihlab/ssblazer/#/datasets/}. The bioinformatics pipeline for the dataset construction of S1 END-seq can be found at \url{https://www.ncbi.nlm.nih.gov/geo/query/acc.cgi?acc=GSM5100382}.

\bibliographystyle{unsrt}  
\bibliography{references}

\end{document}